\documentclass{article}
\pdfoutput=1
\usepackage[english]{babel}
\usepackage[utf8x]{inputenc}
\usepackage[T1]{fontenc}
\usepackage{amsmath,amssymb}
\usepackage{graphicx}
\usepackage[colorinlistoftodos]{todonotes}
\usepackage[colorlinks=true, allcolors=blue]{hyperref}
\usepackage{booktabs}
\usepackage{cite}

\newcommand{\ratio}{f_n/f_p}
\newcommand{\mdm}{M_{DM}}
\newcommand{\UPTC}{\it Escuela de Física, Universidad Pedagógica y Tecnológica de Colombia,\\ \it
Avenida Central del Norte \# 39-115, Tunja, Colombia}

\title{\bf New Constraints on Xenonphobic Dark Matter from DEAP-3600}
\author{Carlos E. Yaguna\footnote{carlos.yaguna@uptc.edu.co}\\ \UPTC}

\date{}

\begin{document}
\maketitle
\begin{abstract}
The first-year results from DEAP-3600, a single-phase liquid argon direct-detection dark matter experiment, were recently reported. At first sight, they seem to provide no new constraints, as the limit lies well within the region already excluded by  three different xenon experiments: LUX, PandaX-II, and XENON1T.  We point out, however, that this conclusion is not necessarily  true, for it is based on the untested assumption that the  dark matter particle couples equally to protons and neutrons. For the more general case   of isosping-violating dark matter, we find that there are regions in the parameter space where DEAP-3600 actually provides the most stringent limits on the dark matter-proton spin-independent cross section.  Such regions correspond to the so-called Xenonphobic dark matter scenario, for which the neutron-to-proton coupling ratio is close to $-0.7$. Our  results seem to  signal the beginning of a new era in which the complementarity among different direct detection targets will  play a crucial role in the determination of the fundamental properties of the dark matter particle.
\end{abstract}

\section{Introduction}
Determining the fundamental nature of the  dark matter particle is one of the most important open problems in particle and astroparticle physics today \cite{Feng:2010gw,Arcadi:2017kky}. In recent years, direct detection experiments have been able to probe new regions of parameter space and to set stringent limits  on the dark matter interactions \cite{Undagoitia:2015gya,Aprile:2018dbl, Akerib:2016vxi, Cui:2017nnn}.  Unfortunately, the  results from direct detection experiments continue to be presented, probably due to historical reasons, in a way that may hinder their true relevance. 

Typically,  direct detection constraints are shown in terms of the  so-called dark matter-nucleon cross section, which differs from  the physically meaningful dark matter-proton cross section. They  would coincide if the dark matter coupled equally to protons and neutrons, but such an assumption is not supported either theoretically or experimentally. In models where the dark matter couples differently to protons and neutrons --the so-called isospin-violating scenario \cite{Kurylov:2003ra, Giuliani:2005my,Chang:2010yk, Kang:2010mh,Feng:2011vu}-- it does not make sense to compare the limits on the dark matter-nucleon cross section obtained from different targets, as is usually done in the literature. Instead, one should compare the limits on the dark matter-proton cross sections, which will depend on the neutron-to-proton coupling ratio of the dark matter particle. As emphasized for example  in \cite{Yaguna:2016bga}, the interpretation of the experimental results may substantially change within this more general isospin-violating scenario.

Recently, the DEAP collaboration reported the results of a dark matter search, based on 231-live days of data taken during the first year of operation, with the DEAP-3600 experiment \cite{Ajaj:2019imk}. DEAP-3600 is a  direct-detection dark matter experiment that uses 3279 kg of liquid argon as target and is located 2km underground at SNOLAB. At first sight, these results seem to provide no new constraints on the dark matter interactions, for the limit on the dark matter-nucleon spin independent cross section lies well inside the region already excluded by three different xenon experiments: LUX \cite{Akerib:2016vxi}, PandaX-II \cite{Cui:2017nnn}, and XENON1T \cite{Aprile:2018dbl}. In this paper, we point out that this conclusion is not entirely correct.  We reexamine these constraints for  isosping-violating dark matter  and find that there are regions in the parameter space where the DEAP-3600 limit is up to a factor of two stronger than that from  XENON1T. These regions, which we delimit and characterize, correspond to the so-called Xenonphobic dark matter scenario \cite{Feng:2013fyw},  for which the neutron-to-proton coupling ratio is close to $-0.7$. Our  results seem to signal the end of the xenon-dominated epoch and the beginning of a new era in which the complementarity among different direct detection targets will be essential in the determination of the fundamental properties of the dark matter particle.

The rest of the manuscript is organized as follows. In the next section, the main features of the isospin-violating scenario are briefly reviewed. Section \ref{sec:results} presents our main results. In it we compare the limits from  DEAP-3600 and  XENON1T for  isospin-violating dark matter. Finally, our conclusions are drawn in section \ref{sec:con}.

\section{Theoretical Framework}
\label{sec:theory}

For completeness, the isospin-violating scenario for dark matter \cite{Kurylov:2003ra, Giuliani:2005my,Chang:2010yk, Kang:2010mh,Feng:2011vu} is briefly reviewed in this section. In particular, we analyze how the interpretation of the experimental results changes within this more general framework. We refer the reader to \cite{Yaguna:2016bga} for a more general discussion and additional references.

The dark matter-nucleus spin-independent cross section, $\sigma_{A_i}$, is generally given by 
\begin{equation}
\label{eq:sigmaM}
\sigma_{A_i}=\frac{4\mu_{A_i}^2}{\pi}\left[f_{p}\,Z+f_{n}\,(A_i-Z)\right]^2\,,
\end{equation}
where $ \mu_{A_i}=\mdm M_{A_i}/(\mdm+M_{A_i})$ is the dark matter-nucleus reduced mass, $A_i$ is the  number of nucleons, and  $Z$ is the nucleus charge. $f_p$ and $f_n$ denote the dark matter coupling to  the proton and the neutron respectively, and are determined by the underlying particle physics model that accounts for the dark matter.  Notice, in particular, that a priori there is no reason  to expect $f_p=f_n$ (It is known, though, that the \emph{assumptions} of Majorana dark matter and minimal flavor violation lead to $f_p\approx f_n$ \cite{Kelso:2017gib}). That is, in general the dark matter particle will couple differently to protons and neutrons. In fact, several models for which $\ratio\neq 1$ have been studied in the literature \cite{Hamaguchi:2014pja,Drozd:2015gda,Frandsen:2011cg,Belanger:2013tla,Martin-Lozano:2015vva,Gao:2011ka,Feng:2013vaa,Okada:2013cba}. Theoretically motivated scenarios with isospin violation include models where the dark matter interactions are mediated by a dark photon, by the $Z^0$ boson, or  by squarks \cite{Kelso:2017gib}. Thus, it is just a historical accident that the peculiar case $\ratio=1$  has become the default dark matter scenario while the more general case $\ratio\neq 1$ is considered special and is referred to as isospin-violating dark matter.

Another relevant quantity is  the dark matter-proton spin-independent cross section, $\sigma_p$, which  is given by
\begin{equation}
\label{eq:sigmap}
\sigma_p=\frac{4\mu_p^2}{\pi}f^2_{p}\,.
\end{equation}
Direct detection experiments, however, typically report their exclusion limits  in terms of the so-called dark matter-nucleon cross section, $\sigma_N^Z$, which can be written as
\begin{equation}
\sigma_N^Z=\sigma_p\frac{\sum_i \eta_i\mu_{A_i}^2\left[Z+(A_i-Z)\ratio\right]^2}{\sum_i \eta_i \mu_{A_i}^2A_i^2}.
\end{equation}
For $f_p=f_n$ (isospin-conservation),  as assumed in most experimental reports, $\sigma_N^Z$ and $\sigma_p$ coincide,  but in general this is not the case. In particular, if $f_p$ and $f_n$ have opposite signs ($\ratio<0$),   one can have destructive interference between the proton and the neutron contributions to the cross section, so that $\sigma_N^Z\ll\sigma_p$.  In general,  it is $\sigma_p$, rather than $\sigma_N^Z$, that is physically meaningful and that should be used to present and compare different experimental results. It is useful, therefore, to define the ratio between these two quantities,
\begin{equation}
F_Z\equiv\frac{\sigma_p}{\sigma_N^Z}.
\end{equation}
$F_Z$, which depends on the target nucleus and on $\ratio$, gives  the factor by which the sensitivity of a direct detection experiment is suppressed for isospin-violating dark matter. In other words, if $\tilde\sigma$ is the limit  on the spin-independent  dark matter-nucleon cross section reported by an experiment (at a given dark matter mass),  then $F_Z \tilde\sigma$ is the  limit on the dark matter-proton cross section that actually applies to the isospin-violating scenario.

\begin{figure*}[tb]
\includegraphics[width=0.9\columnwidth]{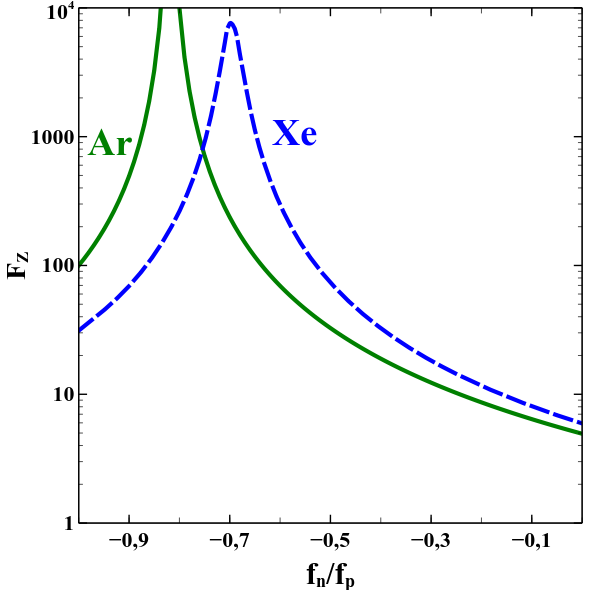}
\caption{The suppression factor, $F_Z$, as a function of $f_n/f_p$ for xenon (blue) and argon (green) targets. $F_Z$ gives the \emph{sensitivity loss} on the dark matter-proton  spin-independent cross section for the case of isospin-violating dark matter. \label{fig:suppression}}
\end{figure*}

Figure \ref{fig:suppression} shows $F_Z$ for xenon (dashed line) and argon (solid line),  and for values of $\ratio$ between $-1$ and $0$.  Because argon consists mostly of a single isotope, there exists a value of $\ratio$ for which there is an exact cancellation between the neutron and proton contributions to the cross section, so that the dark matter does not interact with an argon nucleus. At such point, $\ratio\approx -0.82$, argon experiments completely lose their sensitivity ($F_Z\to \infty$), as shown in the figure. 

On the other hand, since xenon is composed  of several isotopes, an exact cancellation is not possible and $F_Z$ has a maximum value --the relevance of the distribution of isotopes present in each detector was first emphasized in \cite{Feng:2011vu}. This maximum value of $F_Z$ is achieved  for $\ratio\approx -0.7$  and amounts to  about $10^4$.  Xenonphobic dark matter is defined as a dark matter particle featuring a value of $\ratio$ in the vicinity of $-0.7$ \cite{Feng:2013fyw} \footnote{It was called Xenophobic in that work; we have used instead the more precise Xenonphobic.}. Notice that, although highly suppressed, the coupling between a xenon nucleus and a  Xenonphobic dark matter particle is \emph{not} zero. 

\begin{figure*}[tb]
\includegraphics[width=0.9\columnwidth]{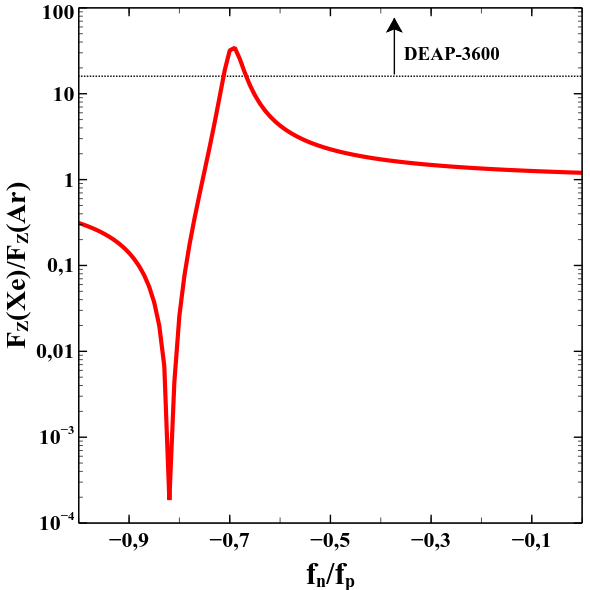}
\caption{The values of $\ratio$  where  DEAP-3600 outperforms  XENON1T for high dark matter masses. Above the dotted black line, DEAP-3600 currently provides the most stringent limit on the dark matter spin-independent cross section.\label{fig:suppressionB}}
\end{figure*}

 According to figure \ref{fig:suppression}, for Xenonphobic dark matter ($\ratio\approx -0.7$),  the limits from xenon experiments (XENON1T, PandaX-II, LUX) are actually weaker by almost four orders of magnitude whereas those from argon experiments by about two orders of magnitude.  Since the gap between the 2018 XENON1T limit  and the recent limit from DEAP-3600 can be  less than two orders of magnitude, the latter may  set the most stringent constraints on Xenonphobic dark matter.  This fact is illustrated in figure \ref{fig:suppressionB}, which shows the ratio between the $F_Z$ factors for xenon and argon (solid line), and compares it against the ratio of the limits from XENON1T and DEAP-3600 at high dark matter masses (dotted horizontal line). Above the dotted line, for $\ratio$ in the Xenonphobic region, DEAP-3600 is expected to set the most stringent limits for large dark matter masses.  In the next section, the region of parameter space where DEAP-3600 outperforms XENON1T is determined.

\section{Results}
\label{sec:results}

\begin{figure*}[tbh]
\includegraphics[width=0.9\columnwidth]{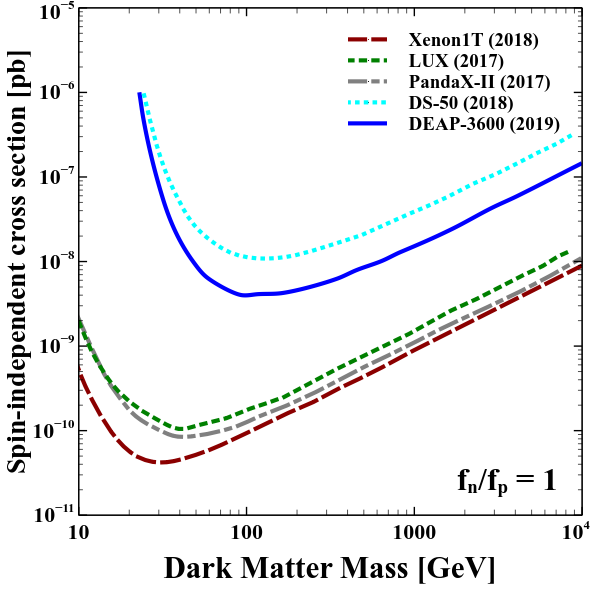}
\caption{Current  limits on the dark matter-proton spin-independent direct detection cross section for the case $f_n/f_p=1$ (isospin-conserving dark matter).  \label{fig:limits}}
\end{figure*}

To begin with, let us illustrate the latest results from DEAP-3600 and how they compare against other experiments under the standard assumption, $\ratio=1$ --see figure \ref{fig:limits}. This figure is quite similar to that shown by the DEAP collaboration in their recent publication \cite{Ajaj:2019imk}. It displays the current limits on the dark matter-nucleon spin-independent cross section from different experiments: DS-50 (dotted line), DEAP-3600 (solid line), LUX (short-dashed line), PandaX-II (dash-dotted line) and XENON1T (dashed line). The upper two lines correspond to argon experiments whereas the lower three lines to xenon experiments. From this figure it seems that the new results from DEAP-3600 are hardly relevant; they are just excluding a region that had already been excluded by three different experiments.  In the following, we will challenge this interpretation. 

The crucial point is that figure \ref{fig:limits} is valid only for  $\ratio=1$, an assumption without any theoretical or experimental support. And as explained in the previous section, the interpretation of these experimental results may drastically change when we consider the more general scenario of isospin-violating dark matter. In fact, we already know that it is for Xenonphobic dark matter ($\ratio\approx -0.7$) that we expect the most significant modifications.

\begin{figure*}[tb]
\includegraphics[width=0.9\columnwidth]{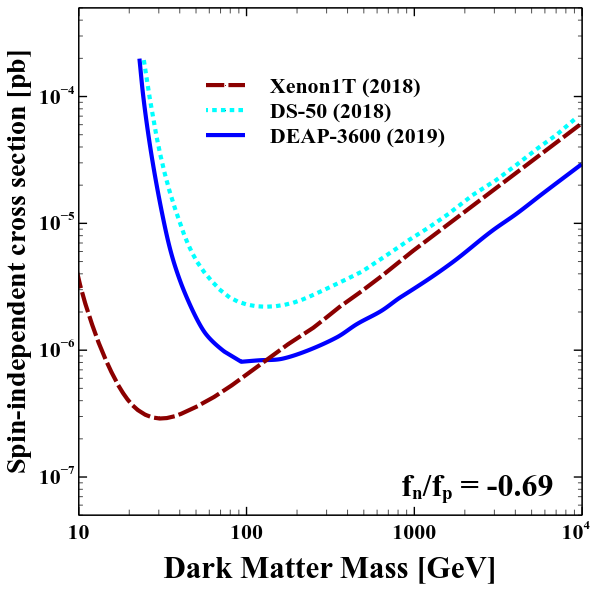}
\caption{Current  limits on the dark matter-proton spin-independent direct detection cross section for the case $f_n/f_p=-0.69$ (isospin-violating dark matter).   \label{fig:limits069}}
\end{figure*}

Figure \ref{fig:limits069} compares the experimental limits from XENON1T, DS-50, and DEAP-3600 for $\ratio=-0.69$. For clarity we dropped from this figure the limits from LUX and PandaX-II, which are always close to and weaker than the XENON1T limit. Notice, first of all, that the scale on the $y$ axis is different from the previous figure, as all limits become weaker --see figure \ref{fig:suppression}. Remarkably, we find that, in this case, the recent limit from DEAP-3600 is actually stronger than that from XENON1T (or any other xenon experiment) in the region $\mdm\gtrsim 130$ GeV.  At high masses, the difference between these two limits amounts to about a factor of two. That is, the recent results from DEAP-3600 are actually probing \emph{new} regions of the parameter space for Xenonphobic dark matter.

This figure not only demonstrates the main thesis of this paper but it also emphasizes the need to find a better way to present and compare the  limits (and future signals) from direct detection experiments using different targets. At the very least, a caveat should be included indicating that the result is valid only for the special case $\ratio=1$.

\begin{figure*}[tb]
\includegraphics[width=0.9\columnwidth]{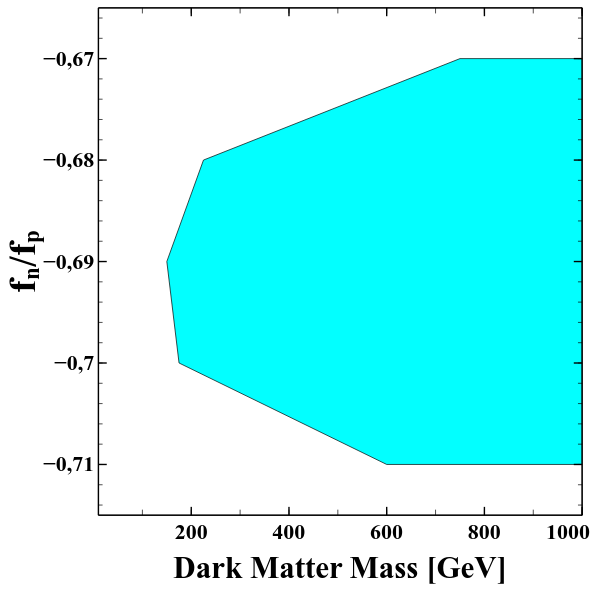}
\caption{The region of parameter space where DEAP-3600 currently sets the most stringent limits on the dark matter-proton spin-independent cross section.  \label{fig:dmfnfp}}
\end{figure*}

Figure \ref{fig:dmfnfp} displays the region of parameter space where DEAP-3600 currently outperforms XENON1T. It includes values of  $\ratio$ between $-0.67$ and $-0.71$, and dark matter masses larger than about $150$ GeV. Given that both experiments are expected to release new limits in the near future, it will be interesting to see how this region gets updated.

In figure \ref{fig:limits069} we also displayed the recent limit from DS-50 (dotted line), another argon experiment, because we wanted to make a point about the qualitative change that has occurred with the first-year limit from DEAP-3600. Notice that the region excluded by DS-50 lies entirely within the XENON1T exclusion region. Thus, DS-50 is not probing new regions of the parameter space. In other words, until the recent limit from DEAP-3600, xenon experiments were so dominant that they were imposing the most stringent constraints even on Xenonphobic dark matter --see also \cite{Yaguna:2016bga}. It is only now, with the release of the first-year of data from DEAP-3600, that this situation has changed. This new development --the fact that argon experiments have already become competitive-- is essential, for it may allow to test, once dark matter signals are observed, if the dark matter interactions are really isospin-conserving ($f_p=f_n$) or not --see e.g. \cite{Kelso:2017gib} for a recent discussion. The observation of dark matter direct detection signals from different targets is also required to test whether the dark matter particle is its own antiparticle, as proposed in \cite{Queiroz:2016sxf,Kavanagh:2017hcl}. Our results seem to indicate that we have just entered into a new era  in which the complementarity among different direct detection targets will play a crucial role in  the determination of the fundamental properties of the dark matter particle.

\section{Conclusions}
\label{sec:con}
We demonstrated that the first-year results from DEAP-3600,  a single-phase liquid argon direct-detection dark matter experiment, are more relevant than they appear at first sight, for they set new limits on Xenonphobic dark matter ($\ratio\approx -0.7$). In fact, the DEAP-3600 limit may exceed by up to a factor of two the current limits from xenon experiments (see figure \ref{fig:limits069}).  In this work, the range of parameter space for which DEAP-3600 sets the most stringent limits on the dark matter-proton spin-independent cross section was delimited (see figure \ref{fig:dmfnfp}).  Finally, we also pointed out that these new limits seem to signal the end of the xenon-dominated epoch and the beginning of a new era in which the complementarity among different direct detection targets will play a crucial role in the determination of the dark matter nature.

\bibliographystyle{hunsrt}
\bibliography{darkmatter}

\end{document}